\documentclass[journal=nalefd,manuscript=letter,layout=twocolumn]{achemso}
\setkeys{acs}{articletitle=true,super=false}
\usepackage[version=3]{mhchem} 
\usepackage[T1]{fontenc}       
\pdfoutput=1
\usepackage{graphpap}
\usepackage{graphicx}
\usepackage{color}
\usepackage{amsmath, amssymb, amsfonts, amsthm}

\newcommand{\onlinecite}[1]{\hspace{-1 ex} \nocite{#1}\citenum{#1}}

\author{Christian~Volk}
\altaffiliation{present address: QuTech and Kavli Institute of Nanoscience, TU Delft, 2600 GA Delft, The Netherlands}
\author{Anasua~Chatterjee}
\author{Fabio~Ansaloni}
\author{Charles~M.~Marcus}
\author{Ferdinand~Kuemmeth}
\affiliation{Center for Quantum Devices, Niels Bohr Institute, University of Copenhagen and Microsoft Quantum Lab Copenhagen, Universitetsparken 5, 2100 Copenhagen, Denmark}
\email{kuemmeth@nbi.dk}

\date{ \today}

\title{Fast charge sensing of Si/SiGe quantum dots via a high-frequency accumulation gate}


\begin{document}
\maketitle
\begin{abstract}
Quantum dot arrays are a versatile platform for the implementation of spin qubits, as high-bandwidth sensor dots can be integrated with single-, double- and triple-dot qubits yielding fast and high-fidelity qubit readout. However, for undoped silicon devices, reflectometry off sensor ohmics suffers from the finite resistivity of the two-dimensional electron gas (2DEG), and alternative readout methods are limited to measuring qubit capacitance, rather than qubit charge. By coupling a surface-mount resonant circuit to the plunger gate of a high-impedance sensor, we realized a fast charge sensing technique that is compatible with resistive 2DEGs. We demonstrate this by acquiring at high speed charge stability diagrams of double- and triple-dot arrays in Si/SiGe heterostructures as well as pulsed-gate single-shot charge and spin readout with integration times as low as 2.4~$\mu$s.
\end{abstract}
	
\vspace{18pt}
The exceptional promise of quantum computation is predicated on scalable hardware that can implement multi-qubit devices as well as efficient methods for qubit readout. In recent years, silicon spin qubits based on electrostatically confined quantum dots (QDs) have been shown to fulfill many of these criteria and are therefore promising building blocks for quantum information applications~\cite{Loss1998,Vandersypen2017}. Due to their low concentration of nuclear-spin-carrying isotopes and established fabrication methods, Si/SiGe heterostructures have particular potential for achieving scalability and fault tolerance~\cite{Zwanenburg2013}. While single-qubit~\cite{Kawakami2014,Yoneda2018} and two-qubit~\cite{Veldhorst2015,Huang2018,Zajac2018,Watson2018} operations have been demonstrated with high fidelities, qubit initialization and measurement times are relatively slow. In contrast, in GaAs QD systems, radio-frequency (RF) reflectometry allows fast measurement of charge states~\cite{Reilly2007}. Single-shot readout of spin states employs spin-to-charge conversion in combination with a capacitively coupled sensor dot or a nearby quantum point contact~\cite{Elzerman2004,Barthel2010}. Typically, one low-resistance ohmic contact of the sensor is wirebonded to a surface-mount inductor, forming a RF tank circuit that sensitively responds to changes in the sensor resistance and, indirectly, to the qubit's spin states~\cite{Barthel2010}.  Singlet and triplet states were distinguished with a signal-to-noise ratio ($SNR$) as high as 6 for integration times as low as 200~ns~\cite{Higg2014}. An application of this technique to accumulation mode silicon devices is possible for carefully designed, high-quality samples~\cite{Takeda2016}, but raises specific challenges in contrast to depletion mode GaAs devices: the strong capacitive coupling of the accumulation gate to the 2DEG changes the matching condition of the resonant circuit significantly and, in conjunction with the relatively large 2DEG resistance, impedes RF readout via the sensor's ohmic contacts. Alternative approaches based on dispersive sensing connect the tank circuit to the plunger gate, such that the reflected RF signal changes when the (quantum) capacitance of the gate electrode changes. This technique, pioneered in GaAs double dots~\cite{Petersson2010} and later applied to silicon devices~\cite{Betz2015,Rossi2016,Crippa2018}, recently allowed single-shot readout of long-lived $T^-$ states~\cite{Urdampilleta2018,Pakkiam2018,West2018}, by decreasing the detection bandwidth to the order of kHz. Replacing the off-chip surface-mount inductor by an on-chip high-impedance superconducting resonator significantly increased the single-shot detection bandwidth (0.3~MHz in Ref. ~\onlinecite{Zheng2019}), but constraints device geometries, materials, and fabrication.

Here, we report high-bandwidth charge sensing compatible with pulsed-gate operation of silicon spin qubits, without the need for nanofabricating additional superconducting elements. We demonstrate this readout technique in undoped Si/SiGe heterostructures, using a single-gate-layer design to form tunable double and triple quantum dot devices. Our reflectometry circuit is galvanically isolated from the heterostructure, by wirebonding a resonating inductor to the accumulation gate of the sensor dot. By decoupling the sensor's ohmic from the RF ground of the sample board, the reflectometry RF signal effectively becomes sensitive to the sensor's conductance, rather than only its capacitance. We thereby achieve single-shot charge and spin readout of proximal quantum dots with integration times on the order of a few microseconds.

\begin{figure}[]
		\centering
		\includegraphics [width=\linewidth] {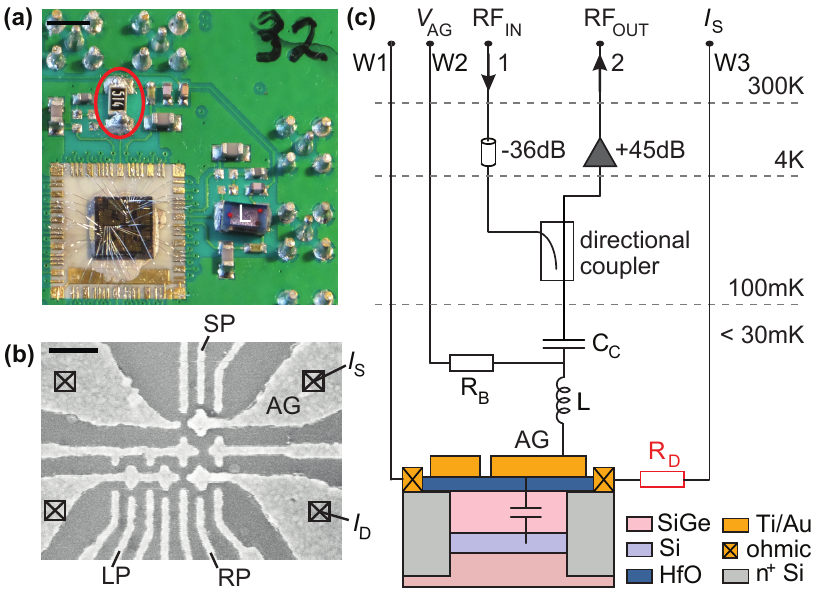}
		\caption[]{
			\textbf{Device design and reflectometry circuit.}
(a) Silicon-germanium chip wirebonded to a PCB-mounted inductor (L), a decoupling resistor $\mathrm{R_D}$ (red circle), and conventional slow and fast signal lines.
Scale bar 3~mm.
(b) Scanning electron micrograph of a representative triple dot (plunger gates LP and RP indicated) with proximal charge sensor (plunger gate SP indicated). The accumulation gate AG is used for reflectometry, whereas four ohmic contacts (crosses) to the 2DEG allow measurements of sensor current ($I_\mathrm{S}$) or device current ($I_\mathrm{D}$).  Scale bar 200~nm.
(c) Simplified reflectometry schematic, showing how a RF carrier applied to the cryostat (port 1) excites the L-AG resonator. The directional coupler allows room-temperature detection of the reflected carrier, with high signal-to-noise ratio due to the use of cryogenic attenuation (-36~dB), amplification (+45dB), and a room-temperature homodyne mixer connected to port 2. The operating voltage of the accumulation gate ($V_\mathrm{AG}$), applied via the $\mathrm{R_B}$-$\mathrm{C_C}$ bias tee, and the resonator's RF voltage capacitively couple to the heterostructure's silicon channel (purple), which is decoupled from the low-pass-filtered cryostat wire (W3) via $\mathrm{R_D}$. Other electrodes and ohmics do not have a decoupling resistor.
		}
		\label{fig1}
	\end{figure}

Our quantum dot devices are fabricated from commercially grown, undoped, natural abundance Si/Si$_{0.7}$Ge$_{0.3}$ heterostructures, schematically shown in Fig.~\ref{fig1}c. Details of the heterostructure, gate fan-out and the fabrication process are provided in the Supporting Information. The Si channel is 42~nm below the gate dielectric, which is grown by atomic layer deposition of HfO$_2$. To avoid a global accumulation gate, a single gate layer patterned by electron-beam lithography defines four large-area accumulation gates and several skinny depletion gates, appropriate to control a triple dot (Fig.~\ref{fig1}c) or double dot (Supporting Fig.~S2d) with proximal  sensor dot. Accumulation gates are operated at positive voltages to accumulate electrons in the 2DEG at the position of the quantum dots, the sensor dot and the source/drain reservoirs. Negative depletion-gate voltages control the electrochemical potential of the dots and thus the electron occupations, as well as the tunnel couplings.
	
All measurements are performed in a dilution refrigerator with electron temperature below 100~mK. The cryostat is equipped with low-pass filtered twisted pairs (DC lines), attenuated semi-rigid coaxial cables (fast-gate lines), reflectometry hardware (see below), and a superconducting magnet. The undoped quantum wells are insulating at cryogenic temperatures. By temporarily illuminating the chip with a red light-emitting diode while applying a negative gate voltage, a carrier density is subsequently induced at relatively small (positive) accumulation voltages. This effect is also observed in etched Hall bar devices fabricated on the same material, where one gate electrode covers the entire active region (Supporting Fig.~S1b). Magnetotransport measurements on such devices confirm the dependence of carrier density on accumulation gate voltage and illumination conditions, and further characterize the quality of the two-dimensional electron gas (see Supporting Fig.~S1c for a measurement of quantum Hall plateaus and Shubnikov-de-Haas oscillations). Low-temperature mobilities of up to 10$^5$~cm$^2$/Vs are achieved at charge carrier densities around $5\cdot10^{11}$cm$^{-2}$. 
The sheet resistivity measures approximately 1.6~k$\Omega/\square$.

A simplified schematic of our reflectometry readout circuit is shown in Fig.~\ref{fig1}c. A surface-mount inductor (L), located on a PCB sample holder (Fig.~\ref{fig1}a), is wirebonded to the accumulation gate (AG) of the sensor. The effective  capacitance associated with the bond wire (which includes stray capacitance in the PCB and, importantly, a capacitive coupling between the accumulation gate and the underlying 2DEG) and the inductance (1200 nH, Coilcraft 1206CS-122XJEB) forms a RF tank circuit. During reflectometry measurements, a RF carrier is applied to the cryostat, and excites the tank circuit via attenuators ($-36$~dB), a directional coupler (-20~dB), and a coupling capacitor ($\mathrm{C_C}$). The tank-circuit response is measured by amplifying the reflected carrier at 4~K (+45~dB, Weinreb CITLF1), followed by homodyne detection at room temperature (Polyphase Microwave quadrature demodulator AD0105B) and sampling of the demodulated voltage ($V_\mathrm{H}$) by a fast digitizing card (AlazarTech ATS9440). To prevent the RF excitation from directly shunting to the RF ground of the sample holder, the sensor ohmic underneath the accumulation gate is connected via a high-impedance decoupling resistor ($R_\mathrm{D} = 0.5$~M$\Omega$) to a DC gate voltage line (W3). The other sensor ohmic is bonded directly to a DC line (W1). A high-impedance bias resistor ($R_\mathrm{B}$) allows the application of a tuning voltage ($V_\mathrm{AG}$) to the accumulation gate.

	\begin{figure}[]
		\centering
		\includegraphics [width=\linewidth] {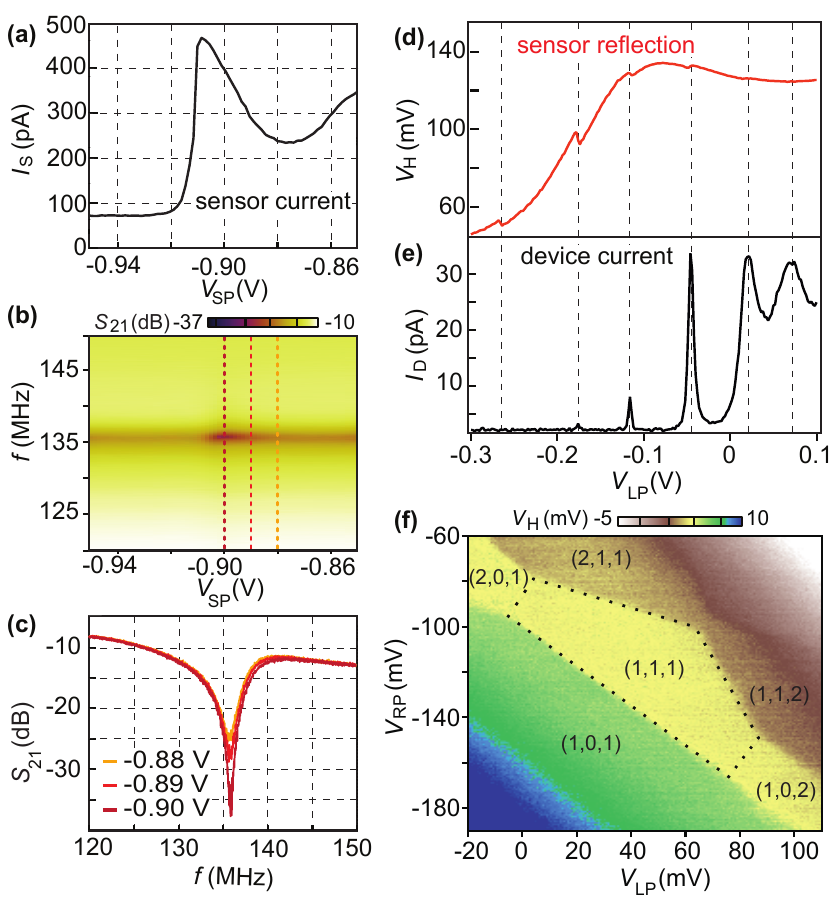}
		\caption[]{
			\textbf{RF charge sensing.}
(a) Sensor dot current as a function of the plunger gate voltage $V_\mathrm{SP}$, for fixed ohmic bias of 500~$\mu$V.
(b) Scattering parameter from port 1 to port 2, $S_{21}$, as a function of $V_\mathrm{SP}$ and carrier frequency $f$.
(c) $S_{21}(f)$ for gate voltages indicated in (b), demonstrating near the 136-MHz resonance a sensitivity of carrier reflection to changes in sensor conductance.
(d) Demodulated voltage $V_\mathrm{H}$ from homodyne detection at 136~MHz, as a function of the left-plunger voltage $V_\mathrm{LP}$. (e) Simultaneous $I_\mathrm{D}$ measurements in the device's Coulomb-oscillations regime indicate that kinks in $V_\mathrm{H}$ result from charging events in the device.
(f) $V_\mathrm{H}$ as a function of the left and right plunger gates, revealing the charge stability diagram of the triple dot device. Single-electron occupation of the three dots is indicated by a dotted line. A plane fit to the central region of (1,1,1) has been subtracted from $V_\mathrm{H}$.
		}
		\label{fig2}
	\end{figure}

Initially, a sensor dot is tuned up in the top half of the device shown in Fig.~\ref{fig1}b, using conventional DC transport measurements via wires W1 and W3. We increase the accumulation gate voltage until a conductive channel is formed, and then operate the barrier gates close to their pinch-off voltage to confine a quantum dot.
Figure~\ref{fig2}a shows a transport measurement of a Coulomb resonance of the sensor dot as a function of the plunger gate voltage. 
(In this configuration, we estimate that the resistance between one of the dot's barriers and the respective wirebonding pads is 20~k$\Omega$, including ohmic contact resistance and finite resistivity of the 2DEG, i.e. a significant fraction of the applied bias voltage drops over the decoupling resistor.) 
Simultaneously, the demodulated voltage of the reflectometry circuit has been measured as a function of the applied RF frequency (Fig.~\ref{fig2}b). The reflected RF power is strongly modulated by the conductance of the sensor dot. The minimum, i.e. when the resonant circuit is matched best, approximately aligns with the Coulomb peak. The resonance frequency stays constant indicating that the capacitive and inductive contributions to the readout circuit are not affected. Fig.~\ref{fig2}c compares cuts through (b) at selected gate voltages, showing a resonance dip at 136~MHz. The reflected power at resonance changes by 12~dB, while the current of the sensor dot changes by 170~pA.

By tuning the sensor dot to the flank of a Coulomb peak, the reflected RF amplitude becomes sensitive to the charge within the triple-dot channel. The RF frequency, power and phase are optimized for best readout contrast. First, we tune up a single QD in the triple-dot channel (bottom half of Fig.~\ref{fig1}b). A measurement of the sensor reflection $V_\mathrm{H}$ as a function of the triple-dot plunger gate voltage is shown in Fig.~\ref{fig2}d. The signal shows steps in amplitude that align well with the Coulomb peaks of the triple-dot device measured simultaneously in DC transport (Fig.~\ref{fig2}e).
In addition, the sensor reflection is sufficiently sensitive to resolve charge transitions in regimes where the DC current through the triple-dot device is below the detection limit (for instance, see the left most charge transition). This is especially relevant for tuning up quantum dot arrays with single-electron occupations, appropriate for many spin qubit experiments. As an example, we tune up a triple QD configuration where each of the QDs is filled with one electron. The charge stability diagram (Fig.~\ref{fig2}e) shows the typical pattern of a triple QD. The demodulated voltage is plotted as a function of the left and right plunger gates, as labelled in Fig.~\ref{fig1}b.

	\begin{figure}[]
		\centering
		\includegraphics [width=\linewidth] {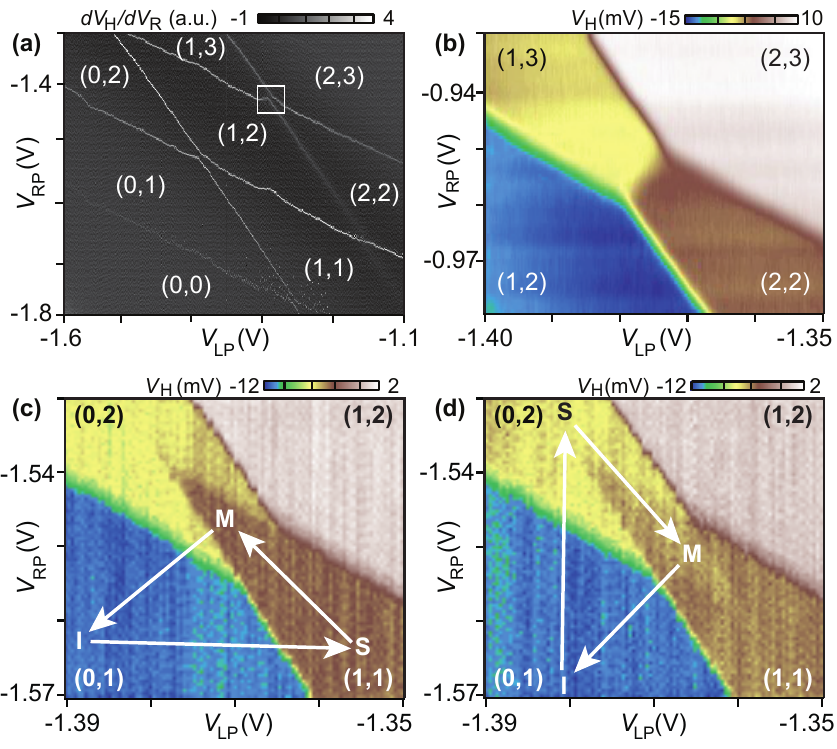}
		\caption[]{\textbf{Interdot charge relaxation in a pulsed-gate double dot.}
(a) Charge stability diagram of a few-electron double dot device (derivative $dV_\mathrm{H}/dV_\mathrm{RP}$ plotted for clarity). Numbers (n,m) indicate occupation of the left and right dot, respectively.
(b) High-resolution zoom on the charge transition highlighted in (a), after tuning. Total acquisition time 1~s.
(c-d) Three-step voltage pulses (arrows) are repeatedly applied to the left and right plunger gate, while slowly changing the DC voltages $V_\mathrm{LP,RP}$ such that $V_\mathrm{H}$ represents the average over many pulse repetitions. The M/I/S segments of the pulse are 5/1/1$~\mu$s long, with the RF carrier applied only during the M segment.
(c) For counterclockwise pulse trajectories, a pulse triangle of (1,1) character appears in the region near "M", indicating that relaxation from Pauli-blocked (1,1) states to the (0,2) ground state exceeds $5\mu$s.
(d) For clockwise pulse trajectories, no reversed pulse triangle is visible, indicating the relaxation between (0,2) and (1,1) occurs at much shorter time scales. For better charge visibility in panels b, c, d, a plane fitted to (1,2) or (2,3) regions has been subtracted from $V_\mathrm{H}$.
		}
		\label{fig3}
	\end{figure}

Next, we demonstrate fast device characterization that takes advantage of the high bandwidth of our reflectometry technique. Figure~\ref{fig3}a shows the charge stability diagram of a double QD in the low-electron regime (the device is shown in Supporting Fig.~S2d). To speed up this acquisition, a 2-kHz saw-tooth pulse is applied to one of the plunger gates while stepping the other. The frequency is chosen to be larger than the cut-off frequency of the bias tee, but smaller than typical tunnel rates to avoid electron latching effects. This technique allows a high-resolution scan of charge stability regions within one second (for example Fig.~\ref{fig3}b shows the (1,2), (2,2), (1,3), and (2,3) ground state regions), compared to acquisition times of several minutes using conventional DC transport measurements. At reduced resolution, video rate scans are possible, which facilitates the measurements significantly, especially allowing a ``real-time'' tuning procedure. The charge stability diagram can then be continuously monitored while adjusting other parameters, such as the tunnel couplings.

Our reflectometry technique also allows pulsed-gate measurements typical of time-domain spin qubit experiments, such as the determination of spin and charge dynamics. In order to determine spin life times directly, nanosecond-to-microsecond-long gate pulses are used, along with spin-to-charge conversion based on Pauli spin blockade, a common readout technique to distinguish between singlet and triplet states~\cite{Johnson2005,Barthel2010,Borselli2011,Urdampilleta2015}. To probe these effects in our devices, we apply a three-step pulse cycle to the plunger gates. First, the double QD is initialized in the (0,1) occupation (position I in Fig.~\ref{fig3}c), followed by a pulse to separation point (S) where an electron of random spin state is loaded from the reservoir. Readout takes place at the measurement point (M), located in the (0,2) ground state region. A (1,1) singlet state can relax into the energetically favorable (0,2) singlet state, whereas a (1,1) triplet state remains in (1,1) until a spin flip takes place, due to Pauli spin blockade. By applying the RF readout tone only during the M step, the resulting (averaged) reflectometry signal distinguishes between the (0,2) and (1,1)  charge states selectively during the M step, and thus provides information about triplet-to-singlet relaxation rates~\cite{Barthel2010}.

In Fig.~\ref{fig3}c, we record a charge stability diagram while repeatedly applying the pulse cycle described above. The brown region extending from the (1,1) ground state region into the (0,2) ground state region (pulse triangle) shows that the system cannot immediately relax into the (0,2) ground state, indicating the presence of Pauli spin blockade. Thus, the duration of the M step ($5\mu$s) gives a lower bound for the spin relaxation time. In Figure~\ref{fig3}d we show a control measurement with an inverted gate pulse trajectory. Here, no such pronounced pulse triangle is visible, in agreement with the expectation that no spin blockade is present in the charge transition from (0,2) to (1,1). Instead, a faint rhombus-shaped region with an average charge between (0,2) and (1,1) appears, likely related to averaging over instrinsic metastabilities within the double dot~\cite{Biesinger2015}.

	\begin{figure}[]
		\centering
		\includegraphics [width=\linewidth] {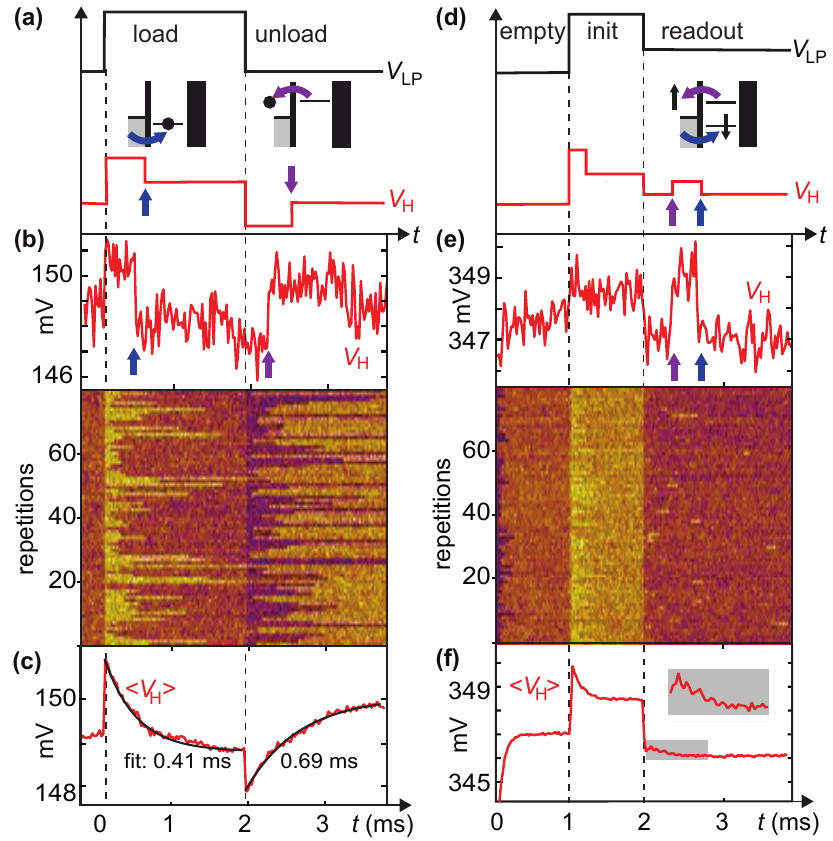}
		\caption[]{
			\textbf{Single-shot readout of a quantum dot.}
(a) Square pulse (black) repeatedly applied to the left plunger gate of a triple dot, pulsing across the 0-1 charge transition of the leftmost QD. Charge sensor response (red) expected for detection of an individual electron tunneling onto or off the dot (arrows).
(b) Single-shot trace $V_\mathrm{H}(t)$ acquired during one representative pulse cycle of (a), along with 80 repetitions (lower panel).
(c) Average of 200 single-shot traces (red) with 1/e time from exponential fit to selected ranges (black).
(d) Three-level pulse (black) for single-shot spin readout~\cite{Elzerman2004} repeatedly applied across the 0-1 charge transition. Expected charge sensor response (red) for a spin-up electron, with arrows indicating the characteristic out-in tunnel event during the readout step. This event is absent for spin-down electrons, provided the two spin states straddle the chemical potential of the left reservoir (gray).
(e) Single-shot trace $V_\mathrm{H}(t)$ acquired during one pulse cycle of (d), along with 80 repetitions (lower panel).
(f) Average of 1000 single-shot traces. The inset highlights the presence of a bump, indicative of an ensemble of spin-up events with stochastically distributed tunneling times.
		}
		\label{fig4}
	\end{figure}

The measurements presented so far were obtained by averaging over multiple pulse cycles. To gain a deeper insight into the dynamics of a system, single-shot measurements are an important technique~\cite{Elzerman2004,Barthel2009,Prance2012}. To show single-shot readout, we apply the RF carrier continuously, and first characterize single-electron charge transitions between a QD and an adjacent reservoir, and focus on spin effects later. For that purpose, we apply a square pulse to the left plunger gate of a triple QD, periodically pulsing the left dot across the 0-1 charge transition to load and unload one electron within each period (see Supporting Fig.~S4a). Figure~\ref{fig4}a illustrates the applied pulse cycle together with the expected response of the charge sensor signal. The electrostatic effect of one electron entering or leaving the QD manifests itself as a step in the demodulated voltage $V_\mathrm{H}$, as indicated by the arrows. In order to not miss transitions, the pulse period needs to be sufficiently long compared to the characteristic tunneling time. Due to unintentional capacitive coupling between the plunger gates of the triple dot and the sensor dot, $V_\mathrm{H}$ also shows steps whenever the plunger voltage changes (black dashed lines).

Figure~\ref{fig4}b shows a representative single-shot readout trace from one such pulse cycle, using a pulse period of 3.6~ms and an integration time of 24~$\mu$s per data point. The arrows highlight the charge sensor response to an electron tunneling in and out from the dot. (Single-shot traces with integration times as small as 2.4~$\mu$s are discussed in Supporting Fig.~S6.) Repeated acquisition of many single-shot traces as in the lower part of Fig.~\ref{fig4}b provide statistics of single-electron tunneling times. For example, the average over 200 single-shot traces is shown in Fig.~\ref{fig4}c, yielding tunnel in (out) times of 0.41 (0.69) ms from exponential fits for this particular tuning. Alternatively, software detection of tunneling events based on wavelet analysis\cite{Prance2015} yields tunnel rates in good agreement with those obtained from the exponential fits (see Supporting Figs.~S5 and S6).

Finally, we apply a pulse cycle designed to detect spin-dependent tunneling from the QD to the reservoir. The spin degeneracy is lifted by an in-plane magnetic field of 800~mT. We apply a three-step pulse cycle consisting of an empty, initialization and readout step~\cite{Elzerman2004}, as illustrated in Fig.~\ref{fig4}d. First, the energy of both spin states is raised above the Fermi level of the reservoir to empty the QD. Then, the initialization step pulses both states below the Fermi level to load an electron of random spin orientation. Subsequently, spin-selective tunneling is achieved if the readout pulse places the Fermi level just between the Zeeman-split spin states of the QD: A spin-down electron will remain on the QD, while a spin-up electron can tunnel out to the reservoir before a spin-down electron repopulates the QD. The characteristic "electron out electron in" tunneling events associated with spin up show up as a temporary change in the sensor response, as illustrated with arrows in Fig.~\ref{fig4}d. Spin-selective tunneling requires the plunger gate voltage in the readout step be chosen correctly, such that spin-split QD states straddle the Fermi level. We tuned to this readout position by repeatedly applying the three-step pulse cycle while slowly stepping the DC gate voltage of the plunger gate until the readout characteristics were observed (see Supporting Fig.~S4). For this procedure to work, the Zeeman splitting ($\approx 90~\mu$eV) needs to exceed the thermal energy ($<10~\mu$eV), a condition which is fulfilled in the experiment.
	
Figure~\ref{fig4}e shows a single-shot trace representative for a spin-up QD, with the readout step beginning at 2~ms. The out-in tunneling events can be clearly seen in the charge sensor response (arrows).
With a rms noise level of 0.42~mV in $V_\mathrm{H}$ and a step height of 2.0~mV, the signal-to-noise ratio associated with a 24-$\mu$s integration time is $SNR=\frac{2.0}{\sqrt{2} \cdot 0.42}=3.4$, corresponding to an effective charge sensitivity of $1.5\cdot10^{-3}e/\sqrt{\mathrm{Hz}}$. Assuming that the power signal-to-noise ratio ($SNR^2$) scales linearly with the integration time, we estimate a minimum integration time $t_\mathrm{min}=2.1~\mu$s to achieve $SNR=1$~\cite{Zheng2019}. The 2D plot shows data for 80 repetitions of the same pulse cycle; as expected, some shots show no in-out tunneling events and some of them do. The analysis of spin-down and spin-up traces can be automated using simple thresholding methods, leading to reliable results only at sufficiently high signal-to-noise ratios. An alternative technique, which has been found to be more robust against low-frequency noise and signal drift, is based on wavelet edge detection~\cite{Prance2015}. An example of such a wavelet analysis is shown in Supporting Fig.~S6. Alternatively, the presence of spin-up occupations shows up as a "spin bump" when averaging over many single-shot traces (see inset to Fig.~\ref{fig4}d), with the shape of the spin bump governed by the tunneling rates~\cite{Hayes2009}.

In this work, we demonstrated a high-frequency single-shot readout technique compatible with multi-quantum-dot spin-qubit devices, which we fabricated via a single-layer gate stack in undoped Si/Si$_{0.7}$Ge$_{0.3}$ heterostructures. By connecting a surface-mount inductor to the accumulation gate of a sensor dot, while decoupling the sensor ohmic from the RF ground of the sample holder, we were able to make the resonant circuit response sensitive to the sensor conductance, rather than only its quantum capacitance. This allows charge stability diagrams to be acquired at high rates, which significantly speeds the tuning of QD arrays and opens the door to automated tuning procedures. We achieve single-shot charge and spin readout at integration times on the order of a few $\mu$s, which makes this technique applicable to spin qubit readout. The presented technique constitutes a viable alternative to single-shot readout based on dispersive gate sensing, which so far has been limited to a few kHz or has required integration with millimeter-scale nanofabricated on-chip superconducting resonators. Finally, we expect that this technique is not limited to Si/SiGe devices and spin qubits, but will also find wider application for other accumulation mode devices, for example silicon MOS or germanium hole quantum dot devices.

\section*{Competing interests}
The authors declare no competing financial interest.

\section*{Acknowledgement}
We thank Michael Shea for help in developing the PCB sample holder for pulsed reflectometry measurements. We are grateful for Frederico Martins and Filip Malinowski for experimental help and fruitful discussions.
This project has received funding from the European Union's Horizon 2020 research and innovation programme under grant agreement MOSQUITO (No. 688539) and under the Marie Sklodowska-Curie Action Spin-NANO (Nanoscale solid-state spin systems in emerging quantum technologies, Grant Agreement No. 676108), and from the European Union through the FP7 ICT collaborative project SiSPIN (No. 323841). A.C. acknowledges support from the EPSRC Doctoral Prize Fellowship and the Yusuf Ali Travel Bursary.

\bibliography{Volk2019bib}
\clearpage

\clearpage

\renewcommand{\thefigure}{S\arabic{figure}}
\setcounter{figure}{0}
\onecolumn


\section{Supporting Information for \\ Fast charge sensing of Si/SiGe quantum dots via a high-frequency accumulation gate}	

Supporting Information is provided on heterostructure characterization, device fabrication, printed circuit boards, finding the operating point for spin-selective readout, determination of the minimum integration time, and wavelet analysis.

\section{Material characterization and device fabrication}	

\begin{figure*}[h]
			\centering
			\includegraphics [width=1.0\linewidth] {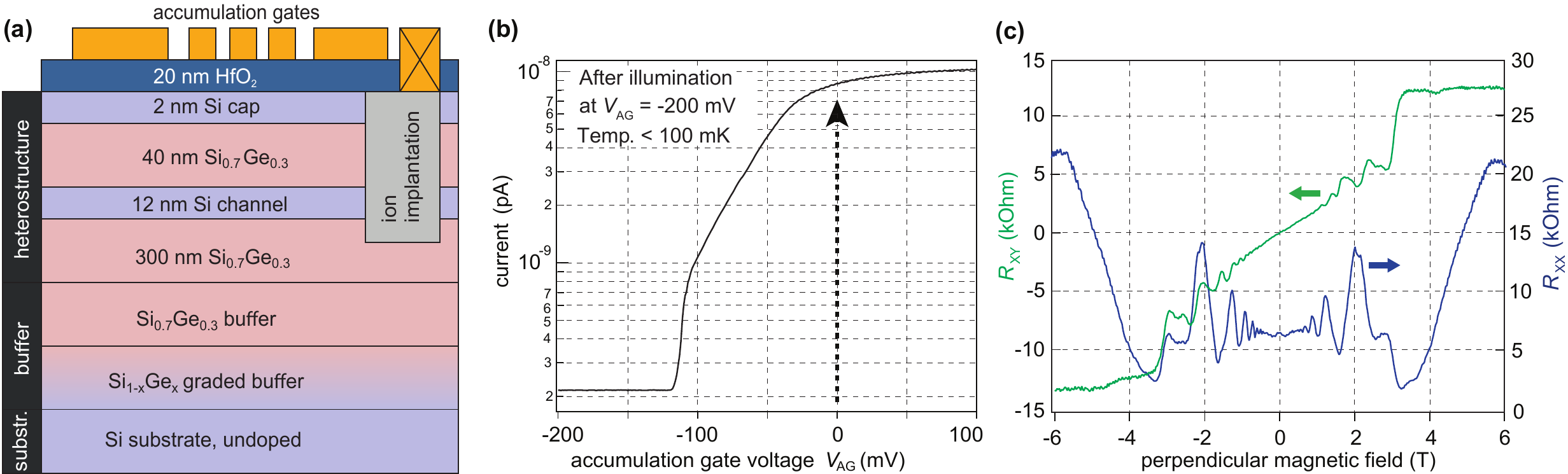}
			\caption[]{\textbf{Material stack and characterization.}
(a) Schematic cross section of the heterostructure grown by chemical vapor deposition on top of a Si substrate: A Si$_{1-x}$Ge$_x$ graded buffer, a 300-nm Si$_{0.7}$Ge$_{0.3}$ layer, a 12-nm Si channel, a 40-nm Si$_{0.7}$Ge$_{0.3}$ layer and a 2-nm Si cap. All layers are undoped with a residual charge carrier density below 10$^{14}$~cm$^{-3}$. HfO$_2$ has been deposited by atomic layer deposition.
(b) Transfer characteristics of a Hall bar. The device has been illuminated at a gate voltage of -200~mV, which sets the turn-on voltage to approximately -110~mV.
(c) Magnetotransport data of a Hall bar. Longitudinal (blue) and Hall (green) resistance, measured at a top gate voltage of +400~mV.
			}
			\label{figS1}
		\end{figure*}

A schematic cross section of the device, including the underlying heterostructure, is shown in Fig.~\ref{figS1}a. A 300-nm-thick Si$_{0.7}$Ge$_{0.3}$ layer is grown on top of a graded buffer in a commercial CVD process, followed by a 12-nm-thick strained Si quantum well, a 40-nm Si$_{0.7}$Ge$_{0.3}$ layer and a 2-nm Si cap (Lawrence Semiconductor Research Laboratory Inc.). 
This places the silicon channel 42~nm below the wafer surface. 
In order to prevent unwanted accumulation and charge leakage, the wafer has been etched outside the device mesa (visible in Fig.~\ref{figS2}c as a 250x250~$\mu$m square), using an Ar$^{+}$-ion milling process that removes the top layers of the wafer including the quantum well. Ohmic contacts are created by phosphorus ion implantation (at energies 30~keV and 15~keV, each at a dose of 1x10$^{15}$~cm$^{-2}$) followed by a 3-minute activation anneal performed at 700$^{o}$C. A layer of HfO$_2$ (typically 20~nm) grown by atomic layer deposition is used as the gate dielectric. The gate electrodes are then patterned in a lift-off process, using a single electron-beam-lithography step followed by electron beam evaporation of 3~nm~Ti and 20~nm~Au.

To perform material characterization, Hall bars were fabricated on the same wafer, following the same fabrication recipe as for the quantum dot devices, and characterized at millikelvin temperatures. Figure \ref{figS1}b shows the transfer characteristics (after the device has been illuminated with a red LED as described in the main text), demonstrating the presence of carriers already at zero accumulation gate voltage. Standard Hall bar measurements are used to determine density and mobility. Figure~\ref{figS1}c shows representative magnetotransport data, in which quantization of the Hall resistance and Shubnikov-de Haas oscillations in the longitudinal resistance are clearly visible.

\section{Mesa and PCB for accumulation-gate-based sensing }	
\begin{figure}[h]
			\centering
			\includegraphics [width=0.5\linewidth] {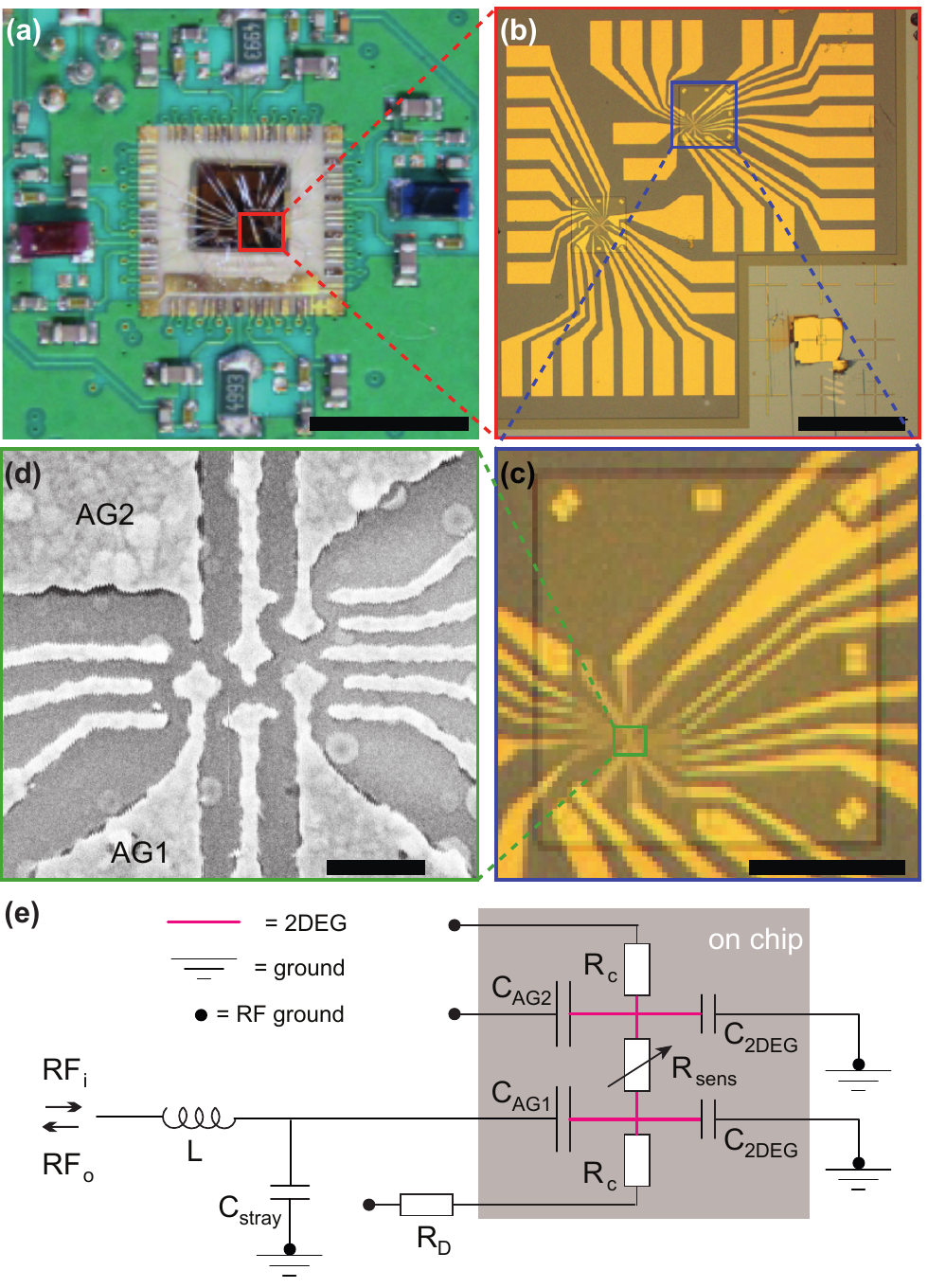}
			\caption[]{\textbf{Quantum dot device mounted for reflectometry measurements.}
				(a) Device chip wirebonded to a PCB sample holder. The chip measures 4x4~mm and hosts eight independent device mesas. Scale bar 8~mm.
				(b) Optical micrograph showing two 250x250~$\mu$m device mesas, each connected to large rectangular wirebonding pads. Scale bar 500~$\mu$m.
				(c) Close-up of one mesa. Near the corners and edges of the mesa, eight regions of ion implantation are visible (gold double squares), which form ohmic contacts to the silicon channel. Scale bar 100~$\mu$m.
				(d) Scanning electron micrograph of the double QD device, showing (in this case) four large-area accumulation gates and 10 skinny depletion gates. Scale bar 200~nm.
                (e) Circuit schematic of the effective RF path on the PCB and on the chip, with RF grounds indicated by dots. The incoming reflectometry signal ($RF_\mathrm{i}$, arriving from the directional coupler, Fig.~1c) reflects off an impedance-matching tank circuit formed by the SMD inductor $L$ and stray capacitance ($C_\mathrm{stray}\approx$~1.2~pF, summarizing contributions from PCB tracks, bond wire and metal tracks on the chip).
                The signal couples via the capacitance of the accumulation gate ($C_\mathrm{AG1} \approx$ 2--5~pF based on geometric estimation) to the underlying 2DEG. The 2DEG has a small unknown capacitance ($C_\mathrm{2DEG}\ll1$~pF) to nearby ground tracks, and a resistive connection to effective RF grounds (black dots) via the sensor quantum dot ($R_\mathrm{sens}\approx0.1$--$0.5$~M$\Omega$) and a contact resistance ($R_\mathrm{C}\approx 20$~$k\Omega$, including contributions from the finite 2DEG resistivity and imperfect ohmic contacts). If the decoupling resistance $R_\mathrm{D}$ is chosen sufficiently high (in our case 0.5 M$\Omega$), and if the admittance $2 \pi f C_\mathrm{AG2}$ is sufficiently high (where $f$ is the carrier frequency and by design $C_\mathrm{AG2} \approx C_\mathrm{AG1}$), then the 2DEG RF excitation reaches RF grounds predominantly via the sensor dot resistance (i.e. the 2DEG part underneath the low-pass filtered acculuation gate, AG2, serves as a RF ground). Overall, this makes the reflected reflectometry signal $RF_\mathrm{o}$ sensitive to changes in $R_\mathrm{sens}$.
			}
			\label{figS2}
		\end{figure}

Figure~\ref{figS2}a shows the central part of the PCB sample holder with a bonded device chip. The LC resonant circuit used for radio-frequency reflectometry measurements is formed by a commercial SMD inductor (Coilcraft 1206CS series) and the stray capacitance associated with the bond wire (approx. $C_\mathrm{stray}$=1.2~pF) connected to the inductor. The bond wire connects to the bonding pad associated with the accumulation gate of the sensor dot, and capacitively couples to the underlying 2DEG (and proximal metallic structures on the chip and PCB).
In the case of Fig.~\ref{figS2}a, two inductors with different values (typically in the range $L$=390-1200 nH) are visible (purple and blue SMD component), which allow frequency-multiplexed readout of multiple sensor dots.
The bonding pad associated with the ohmic contact underlying the accumulation gate is wirebonded to a SMD resistor ($R_D = 0.5$~M$\Omega$, green SMD component), to decouple the RF signal from the RF ground of the sample holder (see caption to Fig.~\ref{figS2}e). The optical micrograph in Fig.~\ref{figS2}b shows a quarter of the chip with two independent devices, each located on its own mesa (one of these mesas is visible in Fig.~\ref{figS2}c as a raised square region).
A close-up of the fine gate electrodes, in this case a sensor dot next to the double quantum dot used for measurements in Fig. 3, can be seen on the scanning electron micrograph in Fig.~\ref{figS2}d.
		
\section{Readout circuit for reflectometry }	
\begin{figure*}[h]
			\centering
			\includegraphics [width=0.9\linewidth] {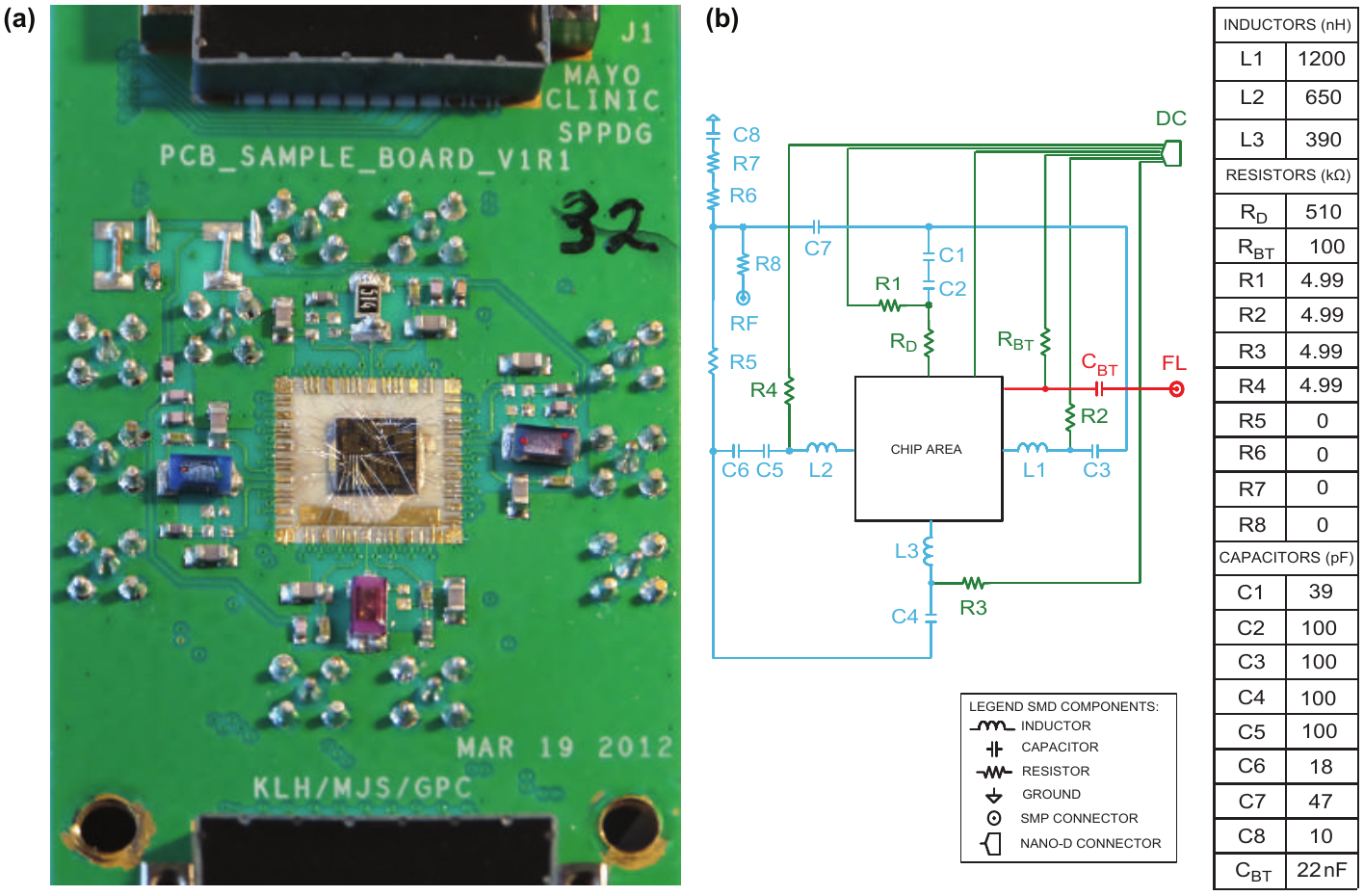}
			\caption[]{\textbf{Configuration of the PCB sample holder.}
(a) Photograph of the PCB sample holder connecting to the cryostat via two low-frequency nanoD connectors (top and bottom) and eleven SMP high-frequency connectors (each mounted from the back side via five through-holes). Three inductors (purple and blue SMDs), one decoupling resistor (black SMD marked 514), as well as some of the bias tees (smaller SMDs) can be identified. Some components are positioned on the back side of the PCB.
(b) Simplified circuit schematic of the PCB, showing signal paths associated with low-frequency control voltages (green), high-frequency control voltages (red), and rf reflectometry signals (blue).
Isolated crossings are achieved by using a multilayer PCB. For clarity, only one high-frequency (low-frequency) bonding pad in red (green) is shown in the upper right corner of the chip area. Symbols are specified in the legend. SMD values are specified in the table.
			}
			\label{figS3}
		\end{figure*}

Figure~\ref{figS3} shows a photograph and a simplified circuit diagram of the printed circuit board (PCB) designed for spin qubit experiments. It comprises 48 DC voltage lines and 10 bias tees (cut-off frequency $\approx$70~Hz). Each bias tee allows the combination of a low-frequency tuning voltage (DC) and a high-frequency manipulation voltage (FL, typically carrying millisecond-to-nanosecond voltage pulses) to be applied to the same wirebonding pad.
The default PCB configuration can be fitted with up to four SMD inductors, such that a single SMP high-frequency connector (RF) is capacitively coupled to up to four resonant LC circuits, each of them equipped with a bias tee to bias and read out up to four charge sensors via frequency multiplexing.
To demonstrate the reflectometry readout method described in the main text, we first modified the circuit by replacing one of the inductors by a decoupling resistor $R_\mathrm{D}$, as shown in Fig.~\ref{figS3}b.
The accumulation gate of the sensor dot is bonded to one of the resonant circuits (L1), and biased via R2, while the ohmic contact underlying the accumulation gate is bonded to the decoupling resistor $R_\mathrm{D}$ (and biased as needed via R1).
For later experiments, we also replaced L3 by a decoupling resistor (as shown in Fig.~\ref{figS2}a), which allows frequency-multiplexed readout of two devices.

\section{Optimization of spin-selective readout point}	

\begin{figure}[h]
			\centering
			\includegraphics [width=0.75\linewidth] {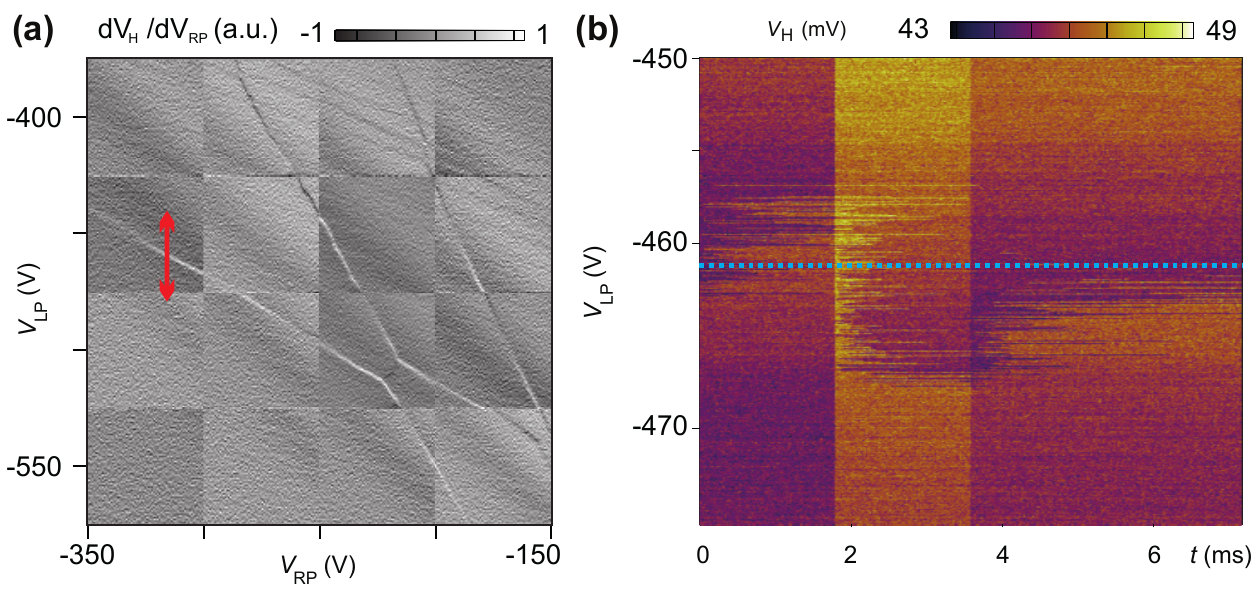}
			\caption[]{\textbf{Readout position for spin-selective readout.}
(a) Charge stability diagram of a device as in Fig.~1b, but tuned up as a double QD. The demodulated reflectometry signal $V_\mathrm{H}$, differentiated with respect to $V_\mathrm{RP}$ is shown for better visibility of charge transitions. The plot has been stitched from sixteen 2D acquisitions, to allow retuning of the sensing dot between acquisitions. The arrow indicates the voltage trajectory of the left plunger gate as the left dot is pulsed across the (0,0)-(1,0) charge transition for single-shot charge and spin readout of Figure~4.
(b) At each DC value of $V_\mathrm{LP}$ the pulse cycle for spin-selective readout is applied (see Fig.~4d of the main text), with each row showing one single-shot readout trace. The dashed line indicates where the readout position is aligned with the charge transition in such a way that the Zeeman-split spin states of the quantum dot straddle the Fermi level of the left reservoir.
			}
			\label{figS4}
\end{figure}

Figure~\ref{figS4} shows supplementary data related to Figure~4 of the main text, describing how the position in gate voltage space for spin-selective readout was found. For these measurements, a triple-dot device is tuned up as a double-dot device. The double dot (0,0)-(1,0) charge transition is first identified using a charge stability diagram (Fig.~\ref{figS4}a). We then apply the pulse cycle for spin-selective readout to the high-frequency connector associated with the left plunger gate, while we step its DC gate voltage (Fig.~\ref{figS4}b). If the gate voltage is far too low ($\lessapprox -468$~mV) or far too high ($\gtrapprox -457$~mV), no tunneling events are observed, indicating that the pulse never crosses the charge addition line and the system remains always either in the (0,0) or (1,0) state. In the range $-468\lessapprox V_{\mathrm{LP}}\lessapprox -462$~mV, the gate voltage is too low and the electron can always tunnel out to the reservoir during the readout step, independent of its spin. In the range $-462\lessapprox V_{\mathrm{LP}}\lessapprox -457$~mV, the electron cannot tunnel out during readout. Only in a small voltage range set by the Zeeman splitting, the spin-split states of the QD straddle the Fermi level of the reservoirs, such that only spin-up electrons can tunnel out from the dot during readout step. This phenomenological procedure was used to determine the readout point for spin-selective readout.

\section{Wavelet edge detection}
\label{wavelet}	

\begin{figure*}[h]
			\centering
			\includegraphics [width=0.9\linewidth] {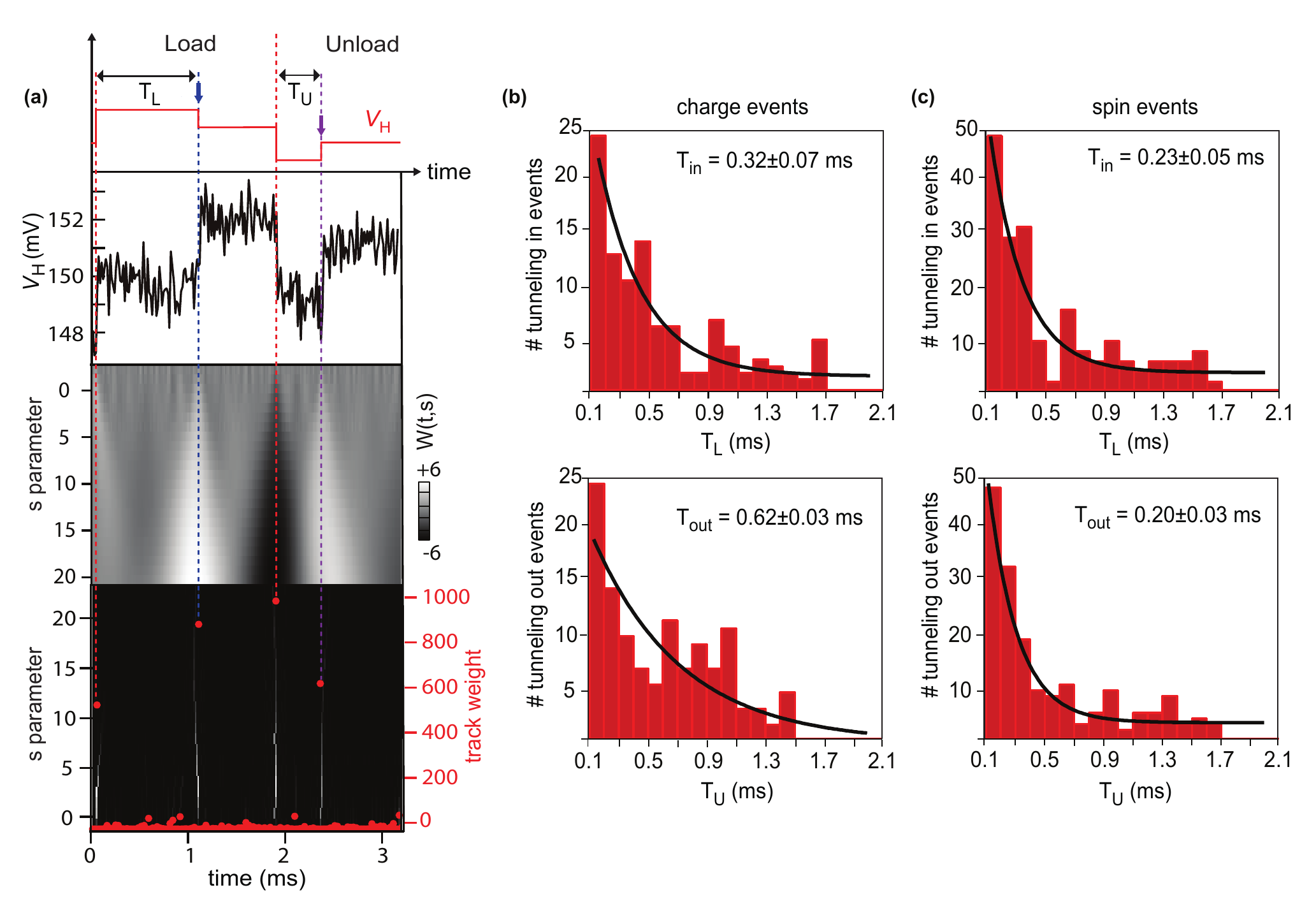}
			\caption[]{\textbf{Wavelet analysis.}
(a) One single-shot trace from the 2D panel in Fig.~4b (black), along with the conceptual definition of event durations for loading ($T_\mathrm{L}$) and unloading ($T_\mathrm{U}$) of an electron. In the presence of noise, tunneling events can be extracted by means of wavelet edge analysis as shown in the lower two panels, based on calculating, weighting, and tresholding tracks using a scaling parameter \textit{s} (see text).
(b) Histogram of the $T_\mathrm{L,U}$ charge events associated with 200 single-shot traces associated with Fig.~4b, extracted using the edge detection algorithm exemplified in (a). Exponential fits (black) yield tunneling times consistent with those obtained from the averaged single-shot traces in Fig.~4c.
(c) Histogram of the $T_\mathrm{L,U}$ spin events associated with 1000 single-shot traces associated with Fig.~4e, extracted by modifying the definitions in (a) appropriate for the spin detection events: $T_\mathrm{U}$  is defined as the time elapsed between the beginning of the readout pulse and the tunnel-out event (purple arrow in Fig.~4d), whereas $T_\mathrm{L}$ corresponds to the time elapsed between the tunnel-out event and the tunnel-in event (blue arrow). Exponential fits (black) yield tunneling times $T_\mathrm{L}$ and $T_\mathrm{U}$ that are approximately identical.
			}
			\label{figS5}
		\end{figure*}

Figure~\ref{figS5} shows the application of a wavelet edge analysis algorithm to data of Fig.~4, allowing automated detection of single-electron-tunneling events as outlined by Prance et al.~\cite{Prance2015}. The technique is based on Canny's edge detection algorithm, used for the recognition of edges in images, and is well suited to detect sharp edges associated with sensor signals. In order to obtain the function \textit{W(t,s)}, the signal $V_\mathrm{H}$ (black trace in ~\ref{figS5}a) is convolved with a scaled mother wavelet, namely the derivative of a Gaussian function of first order, for different scaling factors \textit{s} of the wavelet function. During the second step, shown in the fourth row of ~\ref{figS5}a, the algorithm identifies the track weight for every local minima and maxima at the smallest wavelet scaling factor. The final weight is obtained by summing over the single weights obtained for increasing scaling \textit{s}, for each trace point. The weight is defined as \textit{W(t,s)}$^2$ normalized by the median value of \textit{W(t,s)}$^2$ at a fixed scale.

When the track weight rises above a certain threshold value, here defined as seven times the standard deviation from the average track weight, the event is classified as an edge event. The algorithm is implemented in Igor, with wavelet transformation performed using the Igor CWT function, while the main code is based on MATLAB routines found in the  WaveLab850 library (https://statweb.stanford.edu/\textasciitilde wavelab/). Panels~\ref{figS5}b,c show the tunneling times obtained by applying the wavelet edge detection to the repeated acquisitions presented in the main text (Fig.~4c,f), for charge and spin events respectively.

In order to determine the charge tunneling rates, each single-shot trace is split into two segments, one for each pulse segment. If only one edge is detected within each of these segments, it is recorded as a tunneling event, i.e. either as a loading time ($T_\mathrm{L}$) or unloading time ($T_\mathrm{U}$), depending on whether it occurs in the load or unload segment. The tunneling times are then binned into histograms, using a bin size of 0.1~ms and binning range of 0-2.1~ms (Fig.~\ref{figS5}b).  Fitting exponentials to the histograms (black trace) yields tunneling times consistent with the tunneling times obtained from averaged single-shot traces in Fig.~4c.
The experiment in Fig.~4b was performed at high magnetic field (2~T), suggesting that the difference of tunneling times may either be caused by an accidental (near) degeneracy of two orbitals, or by occupation-dependent and gate-voltage-dependent tunneling barriers.

For the extraction of the spin tunneling times, $T_\mathrm{L}$ or $T_\mathrm{U}$ are defined slightly differently: referring to the pulse cycle of Fig.~4d, $T_\mathrm{U}$  is defined as the time elapsed between the beginning of the readout pulse and the tunnel-out event (purple arrow), whereas $T_\mathrm{U}$ corresponds to the time elapsed between the tunnel-out event and the tunnel-in event (blue arrow). The result of the wavelet analysis is binned to extract the tunneling times only if two edges are detected during the measurement step (Fig.~\ref{figS5}c).
In this case, comparable tunneling times are found for $T_\mathrm{L}$ or $T_\mathrm{U}$, as expected for singly-degenerate levels in the Zeeman-split quantum dot.

Though both experiments were performed for the 0-1 transition of the left dot, we obtained differing transition rates for charge and spin events, possibly due to a small effective shift in tuning voltages and associated tunneling barriers (data in Fig.~4b and 4e were taken several weeks apart). In addition, the rates obtained in this way have a significant uncertainty, which can be improved by increasing the statistics within the histograms.
As a consequence of the conservative thresholding criterion for identifying edge events, only 10\% (2\%) of the single-shot traces were identified with charge (spin) events. This set can likely be increased by optimizing the thresholding criterion.

\begin{figure*}[h]
			\centering
			\includegraphics [width=\linewidth] {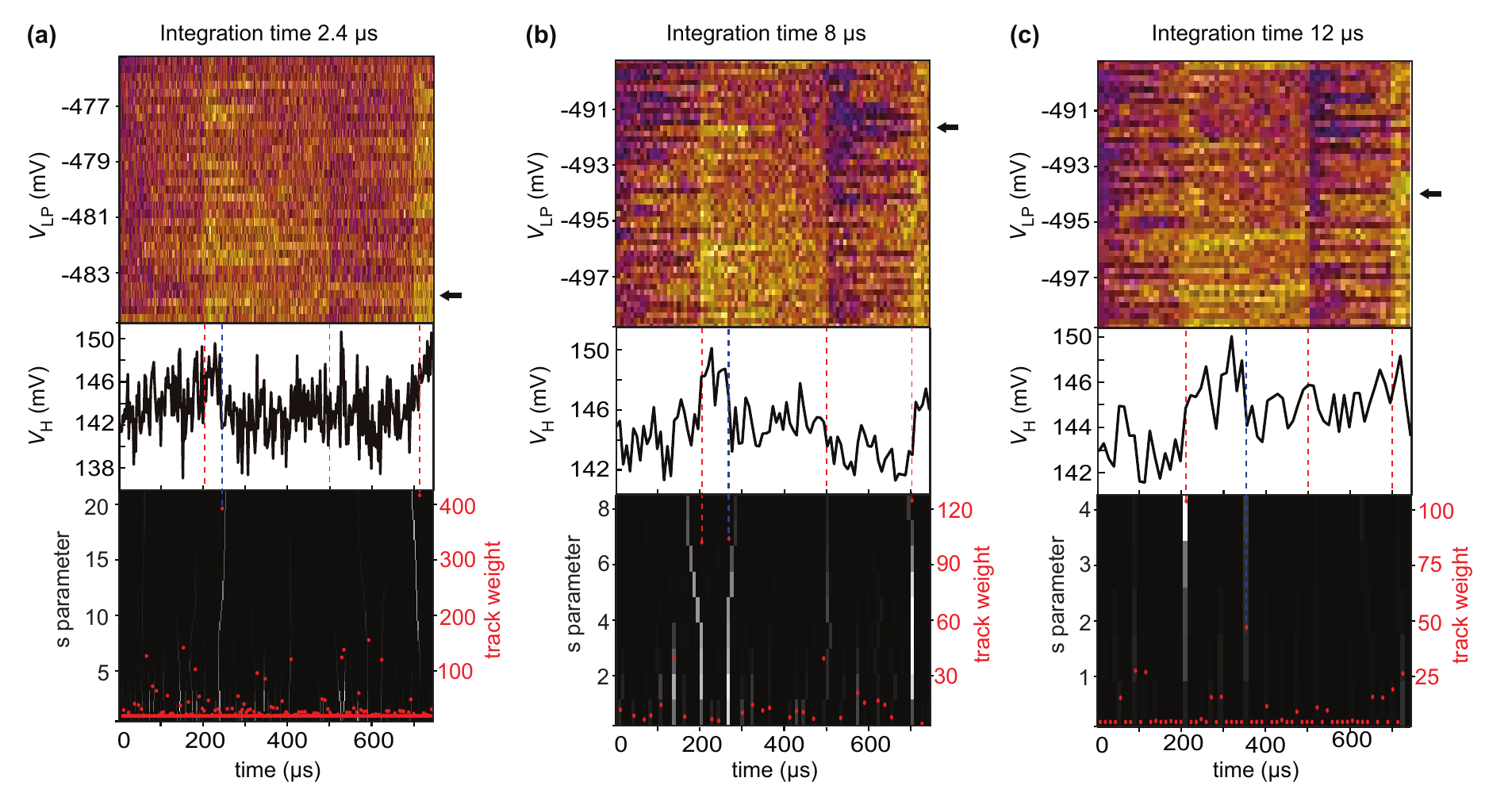}
			\caption[]{\textbf{Wavelet edge detection for noisy single-shot traces.}
				A square pulse (with pulse segments of 200/300~$\mu$s) is repeatedly applied to the left plunger gate to cross the (0,0)-(1,0) charge transition, inducing a cycle similar to that in Fig.~4a. Single-shot traces have been acquired with integration times of 2.4~$\mu$s (a), 8~$\mu$s (b) and 12~$\mu$s (c) per pixel, resulting in ensembles with increasing signal-to-noise ratio. Wavelet edge analysis is used to detect tunneling events into the dot,  exemplified by one representative single-shot trace (black arrow and black trace) for each integration time. Dashed lines mark the sudden variation of $V_\mathrm{H}$ during the acquisition, as detected by the wavelet edge analysis: In red, we mark the steps in $V_\mathrm{H}$ arising from direct capacitive coupling between the left plunger gate and the sensor dot (as discussed in Fig.~4a). In blue, we mark steps due to tunneling events, as identified by a large track weight.
			}
			\label{figS6}
\end{figure*}

To determine the minimum integration time needed to resolve single-electron-tunneling events, we applied a square pulse to repeatedly induce the 0-1 charge transition, using different settings for the integration time associated with the sampling of  single-shot $V_\mathrm{H}$ traces (Fig.~\ref{figS6}). For an integration time as short as 2.4 $\mu$s, tunneling events are hard to detect in the raw data by eye (consistent with our estimation of $SNR\sim$1 for an integration time of 2.1 $\mu$s), yet the wavelet edge analysis still yields useful quantitative results.
		
\end{document}